\documentstyle{nj}
\jl{5}
\def\PII{P_{\! I\! I}}
\def\P34{P_{34}} %P_{\! X\! X\! X\! I\! V}}

\def\be{\begin{equation}}
\def\ee{\end{equation}}

\def\ba{\begin{eqnarray}}
\def\ea{\end{eqnarray}}
\def\ban{\begin{eqnarray*}}
\def\ean{\end{eqnarray*}}

\def\p{Painlev\'e}
\def\bk{B\"acklund}
\def\ep{\varepsilon}
\def\d{{\rm d}}
\def\tfr#1#2{{\textstyle\frac{#1}{#2}}}

\def\^#1{{\bf\widehat{\mit#1}}}
\def\~#1{{\bf\widetilde{\mit#1}}}
\def\L{{\cal L}}
\def\a{\~\alpha}

\def\key#1{\vspace{\baselineskip}
  \noindent\rm Key words: #1\par}

\begin{document}

\title{\bk\ transformations for the %$\mathbf P_{\! I\! I}$
second \p\ hierarchy: a modified truncation approach}[\bk\ transformations
for the $\PII$ hierarchy]

\author{Peter A Clarkson,$^\dagger$ Nalini Joshi$^\ddagger$ and Andrew
Pickering$^{\ddagger,*}$}
\address{$^\dagger$
   Institute of Mathematics and Statistics \\
   University of Kent at Canterbury \\ Canterbury, Kent \\
   CT2 7NF, U.K. \\ \\
   $^\ddagger$
   Department of Pure Mathematics \\
   University of Adelaide \\
   Adelaide 5005 \\
   Australia}

\date{}

\begin{abstract}

The second \p\ hierarchy is defined as the hierarchy of ordinary differential
equations obtained by similarity reduction from the modified Korteweg-de
Vries hierarchy. Its first member is the well-known second \p\
equation, $\PII$.

In this paper we use this hierarchy in order to illustrate our application of
the truncation procedure in \p\ analysis to ordinary differential
equations. We extend these techniques in order to derive auto-\bk\
transformations for the second \p\ hierarchy. We also derive a number of other
\bk\ transformations, including a \bk\ transformation onto a
hierarchy of $\P34$ equations, and a little known
\bk\ transformation for $\PII$ itself.

We then use our results on \bk\ transformations to obtain, for each
member of the $\PII$ hierarchy, a sequence of special integrals.

\end{abstract}

\key{\bk\ transformations, Painlev\'{e} equations}

\pacs{0230, 0340K }

\submitted

\maketitle
\vspace{5truecm}
\hrule\ \\
\noindent$^*${\small Current address: School of Mathematics
and Statistics F07, University of Sydney, Sydney 2006, Australia}\ \\
\vspace{-1truecm}\ \\ \ \\

\newpage

\section{Introduction}
At the turn of the century there was an interest in finding new functions
defined by differential
equations; this led to the discovery of the six \p\ equations
\cite{P00,P02,Gam}, $P_{\! I}-P_{V\!
I}$. These six ordinary differential equations (ODEs) were discovered
through a classification of
second order ODEs of the form
\[ V''(X)=F(V'(X),V(X),X),\qquad '\equiv{\d}/{\d X} \]
where $F$ is rational in $V'(X)$, algebraic in $V(X)$ and analytic in $X$,
whose solutions have no
movable branch points. Here a singularity is said to be movable if its
location in the complex
plane depends on constants of integration. The requirement made was thus
one of single-valuedness
of solutions except at fixed singularities; this property is today referred
to as the \p\ property.

In addition to defining new transcendental functions, the six \p\ equations
have a number of other
remarkable properties. For example, they can each be written as the
compatibility condition of a
corresponding linear system, which then allows them to be solved using the
Inverse Monodromy
Transform Method \cite{FN}. Also, at least for $\PII$--$P_{V\! I}$, they
have a variety of \bk\
transformations and special integrals (see, for example,
\cite{AC91,HA,AMM,BCH,FA,Gr,refIN,MCB} and
references therein). It is the derivation of these last two properties, but
for a hierarchy of
higher order ODEs, that we are interested in here.

As an example, consider the second \p\ equation, $\PII$,
\be
\PII[V,\alpha]\equiv V''-2V^3-VX-\alpha=0.
\label{p2}
\ee
This has the well-known \bk\ transformation
\be
V = \~V +\frac{2\a-\ep}{X+2\~V^2-2\ep\~V'},
\qquad %\label{P2BTa} \\
\alpha = \ep - \a
\label{P2BTb}
\ee
where $\ep=\pm 1$, and where $\~V$ satisfies $\PII$ with parameter
$\a$, that is
\[
\PII[\~V,\a]\equiv \~V''-2\~V^3-\~VX-\a=0.
\]

The second \p\ equation $\PII$ (\ref{p2}) also has special integrals for
certain choices of the
parameter $\alpha$. For example, with $\alpha=\tfr12$ we have the special
integral \cite{Gam}
\be
I_{1/2} \equiv V'-V^2-\tfr{1}{2}X = 0
\label{I11}
\ee
which, modulo $\PII$ with $\alpha=\tfr12$, satisfies the relation
\[
\left(\frac{\d}{\d X}+2V\right)I_{1/2}=0.
\]
The Riccati equation (\ref{I11}) can be linearised via $V=-\psi'/\psi$ to yield
\be
\psi''+\tfr{1}{2}X\psi=0,
\label{peq1}
\ee
which thus gives the well-known Airy function solutions of $\PII$ \cite{Gam}.

The \bk\ transformation (\ref{P2BTb}), together with the discrete symmetry
$(V,\alpha)
\rightarrow (-V,-\alpha)$, then allows us to derive special integrals
$I_{\alpha}$ with
$\alpha=N+\tfr12$, for any integer $N$. Similarly, rational solutions can
be constructed for
integer values of $\alpha$, beginning with the trivial solution $V=0$ for
$\alpha=0$. Later we will
see a little known connection between the sequence of special integrals
(Airy function
solutions), and this sequence of rational solutions.

Modern-day interest in the \p\ equations was to a large extent sparked by
the observation of
Ablowitz and Segur \cite{AS77} (see also \cite{ARS78,ARS8a,ARS8b}) that
they arise as similarity
reductions of many completely integrable partial differential equations
(PDEs). The generalised
symmetries of such PDEs then give immediately a means of writing down
higher order ODEs that may
define new transcendental functions. These form hierarchies of ODEs
obtained as similarity
reductions of hierarchies of completely integrable PDEs. However, although
examples have been
written down (see for example \cite{FN,HA}), it is only recently that much
interest has been shown
in such ODEs \cite{NK97}. The question of whether some of the higher order
ODEs obtained in this
way define new functions remains to be answered. Answering this question is
made more difficult by
the fact that since the time of \p\ only restricted progress has been made
in the
systematic (order by order) classification of ODEs having the \p\ property.

We do not address this particular question in the present work. We are
interested instead in the
question of how to obtain \bk\ transformations and special integrals for
hierarchies of
ODEs. Here we consider one particular hierarchy, the $\PII$ hierarchy
\cite{FN,HA,NK97}, which is
obtained by similarity reduction from the modified Korteweg-de Vries (mKdV)
hierarchy. We show
that the so-called \p\ truncation method, a technique commonly
applied to PDEs
\cite{WTC83,W83}, cannot be applied directly to ODEs. We show how these
difficulties can be
overcome, and accordingly modify the PDE approach in order to obtain
systematically the
auto-\bk\ transformations presented (without derivation) in \cite{HA};
these results provide
a generalisation of (\ref{P2BTb}) above. Using these results we are able to
find sequences of
special integrals for the $\PII$ hierarchy. Such special integrals have not
been presented before
for higher members of this hierarchy. We also give a \bk\ transformation
relating the
$\PII$ hierarchy to another hierarchy of ODEs, namely a $\P34$ hierarchy (so
called because its first
member is equivalent to equation XXXIV of Chapter 14 in \cite{Ince}, which
is commonly referred to
as $\P34$). In addition we show how to obtain a little known \bk\
transformation (due to Gambier \cite{Gam}) for $\PII$
itself. This last allows
us to relate solutions of $\PII$ with parameter value $\alpha=\tfr12$ to
solutions of $\PII$ with
parameter value $\alpha=0$.

\section{The $\mathbf P_{\! I\! I}$ hierarchy}
The Korteweg-de Vries (KdV) hierarchy can be written as
\be
U_{t_{2n+1}} +\partial_x \L_{n+1}[U] = 0, \qquad{} n=0,1,2,\ldots,
\label{kdv}
\ee
where $\partial_x = \partial/\partial x$, and the sequence $\L_n$ satisfies the
Lenard recursion relation \cite{Lax}
\[
\partial_x \L_{n+1} = \left(\partial_x^3+4U\partial_x+2U_x\right)\L_n.
\]
Beginning with $\L_0[U] = \tfr12$, this then gives
\[
\L_1[U] = U, \qquad \L_2[U] = U_{xx} +3 U^2, \qquad
\L_3[U] = U_{xxxx} +10 U U_{xx} +5 U_x^2 +10 U^3,
\]
and so on. The mKdV hierarchy is obtained from the KdV hierarchy via the
Miura map
$U=W_x-W^2$, and can be written as
\be
W_{t_{2n+1}}+\partial_x\left(\partial_x+2W\right)\L_n[W_x-W^2]=0, \qquad{}
n=0,1,2,\ldots
\label{mKdVh}
\ee
The $\PII$ hierarchy \cite{FN,HA,NK97} is obtained from this equation via
the similarity
reduction
\[
W = \frac{V(X)}{[(2n+1)t_{2n+1}]^{{1}/(2n+1)}}, \qquad
X = \frac{x}{[(2n+1)t_{2n+1}]^{{1}/(2n+1)}},
\]
which gives
\be
\PII^{(n)}[V,\alpha_n]\equiv
\left(\frac{\d}{\d X}+2V\right)\L_n[V'-V^2]-VX-\alpha_n=0, \qquad{}
n=1,2,3,\ldots
\label{p2h}
\ee
where we have excluded the trivial case $n=0$ from consideration. We note for
future reference that the $\PII$ hierarchy has the discrete symmetry
$(V,\alpha_n)
\rightarrow (-V, -\alpha_n)$, inherited from the associated discrete symmetry
$W \rightarrow -W$ of the mKdV hierarchy.

Since $\L_1[U]=U$ we see that the first member of this hierarchy is the second
\p\ equation (\ref{p2}). The next member of this hierarchy
is the fourth order equation
\be
\PII^{(2)}[V,\alpha_2] \equiv
V''''-10V^2V''-10V\left(V'\right)^2+6V^5-XV-\alpha_2=0.
\label{p24}
\ee
In what follows, our results on \bk\ transformations and special integrals
will all be presented in general, i.e.\ for any member of the $\PII$
hierarchy. However,
in order to provide concrete examples, we will also give specific results in
the
special cases $n=1$, i.e.\ $\PII$ (\ref{p2}), and $n=2$, i.e.\ $\PII^{(2)}$
(\ref{p24}).

\section{\bk\ transformation for the $\mathbf P_{\! I\! I}$ hierarchy}

\subsection{\p\ truncation for the $\mathbf P_{\! I\! I}$ hierarchy}
For PDEs important information can be obtained using a so-called truncated \p\
expansion \cite{WTC83,W83}. Here we consider the application of this technique
to ODEs, and the modifications that must be made in order to extract
information
about \bk\ transformations. We take as our example the $\PII$ hierarchy
(\ref{p2h}). We begin by making the change of variables
\be
V=\frac{1}{2}\frac{\sigma''}{\sigma'},
\label{str}
\ee
which gives
\be
\left(\frac{\d}{\d X}+\frac{\sigma''}{\sigma'}\right)\L_n\left[\tfr12
S(\sigma)\right]
-\frac{1}{2}\frac{\sigma''}{\sigma'}X-\alpha_n=0, \qquad{} n=1,2,\ldots
\label{se}
\ee
where
\[
S(\sigma)=\frac{\d}{\d X}\left(\frac{\sigma''}{\sigma'}\right)
 -\frac{1}{2}\left(\frac{\sigma''}{\sigma'}\right)^2
\]
is the Schwarzian derivative of $\sigma$. Since the Schwarzian derivative
is invariant under the action of the M\"obius group, setting
$\sigma=-1/\varphi$ gives
\be
V=-\frac{\varphi'}{\varphi}+\frac{1}{2}\frac{\varphi''}{\varphi'}
\label{vphi}
\ee
which is a solution of the $\PII$ hierarchy (\ref{p2h}) provided that
\be
\left(\frac{\d}{\d
X}+\frac{\varphi''}{\varphi'}-2\frac{\varphi'}{\varphi}\right)
\L_n\left[\tfr{1}{2}S(\varphi)\right]
-\frac{1}{2}\left(\frac{\varphi''}{\varphi'}
-2\frac{\varphi'}{\varphi}\right)X-\alpha_n=0, \qquad{} n=1,2,\ldots
\label{e19}
\ee
It then follows that $V$ given by
\be
V=-\frac{\varphi'}{\varphi}+\~V,
\label{tr}
\ee
where
\be
\~V = \frac{1}{2}\frac{\varphi''}{\varphi'},
\label{vt}
\ee
is a solution of the $\PII$ hierarchy (\ref{p2h}) provided that
\be
\left(\frac{\d}{\d X}+2\~V -2\frac{\varphi'}{\varphi}\right)
\L_n[\~V'-\~V^2] -\left(\~V-\frac{\varphi'}{\varphi}\right)X-\alpha_n=0,
\qquad{} n=1,2,\ldots \label{ptr}
\ee
holds.

Equation (\ref{ptr}), or equivalently equation (\ref{e19}), is the result
of substituting
a truncated \p\ expansion into the $\PII$ hierarchy. The \p\
expansion for the principal family $V=-\varphi'/\varphi +\cdots$ has resonances
at $-1$, at $2,3,4,\ldots,2n-2,2n-1$, and at $2n+2$ (these are the same as
those
for the corresponding family of the mKdV hierarchy \cite{FP92}, with the
omission of
that at $2n+1$, which arises from the extra differentiation of the dominant
terms
in (\ref{mKdVh})). Substitution of the truncated expansion (\ref{tr}) into the
$\PII$ hierarchy therefore leads at $\varphi^{-2n}$ to the determination of
$\~V$ as given by (\ref{vt}), and then at $\varphi^{-2n+1}, \ldots,
\varphi^{-2}$ we find that all coefficients vanish since these correspond to
the resonances $2,3,4,\ldots,2n-2,2n-1$. We are then left with the terms at
$\varphi^{-1}$ and $\varphi^0$, and it is these terms that are given by
(\ref{ptr}).

Setting coefficients of different powers of $\varphi$ to zero independently
then gives
\be
\L_n[\~V'-\~V^2]-\tfr{1}{2}X=0
\label{3a}
\ee
and
\be
\left(\frac{\d}{\d X}+2\~V\right)\L_n[\~V'-\~V^2]-\~V X -\alpha_n=0.
\label{3b}
\ee
Together these two equations imply that we must have $\alpha_n=\tfr12$. The
result
of using a truncated \p\ expansion is therefore that if $\~V$ satisfies
(\ref{3a}), then $V$ given by (\ref{tr}) and (\ref{vt}) is a solution of
(\ref{p2h})
for $\alpha_n=\tfr12$. That is, we obtain a map from (\ref{3a}) to
(\ref{p2h}), though
not an auto-\bk\ transformation for the $\PII$ hierarchy.

We do, however, obtain information about special integrals of the $\PII$
hierarchy.
Since $\~V'-\~V^2=V'-V^2$, our result is that any solution of
\[
\L_n[V'-V^2]-\tfr{1}{2}X=0
\]
provides a solution of (\ref{p2h}) for $\alpha_n=\tfr12$. Therefore this
defines
the special integral $I^{(n)}_{1/2}$,
\be
I^{(n)}_{1/2}\equiv \L_n[V'-V^2]-\tfr{1}{2}X=0.
\label{In1}
\ee
This can also be seen from the fact that the $\PII$ hierarchy (\ref{p2h})
can be
written
\[
\left(\frac{\d}{\d X}+2V\right)\left(\L_n[V'-V^2]-\tfr{1}{2}X\right)
+\left(\tfr{1}{2}-\alpha_n\right)=0.
\]

{From} the point of view of the truncated \p\ expansion, since
\be
\~V'-\~V^2=\tfr{1}{2}S(\varphi)=\tfr{1}{2}S(\sigma),
\ee
our result is that if $\varphi$ satisfies
\[
\L_n\left[\tfr{1}{2}S(\varphi)\right]-\tfr{1}{2}X=0,
\]
or equivalently $\sigma$ satisfies
\[
\L_n\left[\tfr{1}{2}S(\sigma)\right]-\tfr{1}{2}X=0,
\]
then
\[
V = -\frac{\d}{\d X}\left(\log\left[\varphi(\varphi')^{-1/2}\right]\right)
= -\frac{\d}{\d X}(\log\sigma')
\]
is a solution of the $\PII$ hierarchy for $\alpha_n=\tfr12$. The change of
variables
\[
\sigma'=\frac{\varphi'}{\varphi^2}=\frac{1}{\psi^2}
\]
then gives that
\[
V=-\frac{\psi'}{\psi}
\]
is a solution of (\ref{p2h}) for any (nonzero) $\psi$ a solution of
\be
\L_n[-{\psi''}/{\psi}]-\tfr{1}{2}X=0.
\label{peq}
\ee

In the special case $n=1$, (\ref{In1}) becomes (\ref{I11}), and (\ref{peq})
becomes (\ref{peq1}). For $n=2$ we obtain the special integral
\be
I^{(2)}_{1/2}\equiv V'''-2VV''+(V')^2-6V^2V'+3V^4-\tfr{1}{2}X
\label{I21}
\ee
which provides solutions of (\ref{p24}) when $\alpha_2=\tfr12$.

We note that in the case $n=1$, this result that the truncated \p\ expansion
leads only to special integrals of $\PII$ appears in \cite{GNTZ}. Weiss,
again for
$n=1$, attempted to overcome this problem \cite{W84} by considering the
symmetries
of a particular integral (the integration constant is set equal to zero) of
equation
(\ref{se}). In this way he was able to obtain, albeit rather implicitly, the
auto-\bk\ transformation for $\PII$. In what follows we give an alternative and
altogether much more explicit derivation of the auto-\bk\ transformations of
the entire $\PII$ hierarchy.

\subsection{Our approach: derivation of the \bk\ transformation}
Thus far we have obtained the special integrals (\ref{In1}) of the $\PII$
hierarchy.
Special integrals in the case $\alpha_n=-\tfr12$ are obtained using the
discrete symmetry
$(V,\alpha_n) \rightarrow (-V, -\alpha_n)$. However our interest is in finding
auto-\bk\ transformations for the $\PII$ hierarchy. It is clear that in order
to do so we need to modify the above truncation approach.

The first point to notice is that equations (\ref{p2h}) and (\ref{3b}) are
copies of
the $\PII$ hierarchy in $V$ and $\~V$, but with the {\it same} value of the
parameter
$\alpha_n$. Yet it is known that auto-\bk\ transformations for ODEs, in
particular the \p\
equations, may involve changes in the value of any parameters in the ODE.
Our first step
therefore is to assume that $\~V$ is a solution of the $\PII$ hierarchy for a
{\it different} choice of parameter, that is
\be
\PII^{(n)}[\~V,\a_n]\equiv
\left(\frac{\d}{\d X}+2\~V\right)\L_n[\~V'-\~V^2]-\~VX-\a_n=0,
\qquad{} n=1,2,\ldots.
\label{p2A}
\ee

We now return to equation (\ref{ptr}). Our added assumption that $\~V$
satisfies
(\ref{p2A}) means that we now have a \bk\ transformation (\ref{tr}) from the
$\PII$ hierarchy in $(\~V, \a_n)$, (\ref{p2A}), to the $\PII$ hierarchy in
$(V, \alpha_n)$, (\ref{p2h}), provided that (\ref{vt}) and
\be
\left(2\L_n[\~V'-\~V^2]-X\right)\frac{\varphi'}{\varphi}
+(\alpha_n-\a_n)=0, \qquad{} n=1,2,\ldots
\label{rh}
\ee
hold. We note now that we cannot set all coefficients of different powers
of $\varphi$ equal to zero in (\ref{rh}), since this would force
$\alpha_n=\a_n$ and we would be
back where we were before, i.e.\ with the standard \p\ truncation. What we
do therefore is
to use (\ref{rh}) to eliminate $\varphi$ from our \bk\ transformation. This is
the most important difference between our approach and that usually used
for PDEs. Our truncated \p\ expansion can then be rewritten in terms of $\~V$,
a second solution of our ODE, only. For PDEs emphasis is usually placed on
the singular
manifold equation \cite{W83,MC91,P96}, but we do not actually solve for
$\varphi$ as
we do here.

The compatibility of (\ref{rh}) with (\ref{vt}) requires
\[
\left(\frac{\d}{\d X}+2\~V\right)\L_n[\~V'-\~V^2]-\~VX
+\tfr{1}{2}(\alpha_n-\a_n-1)=0,
\qquad{} n=1,2,\ldots,
\]
and this last is consistent with (\ref{p2A}) provided that
\[
\alpha_n=1-\a_n.
\]
Using (\ref{rh}), we can now rewrite our truncated \p\ expansion (\ref{tr}) as
\[
V=\~V +\frac{2\a_n-1}{X-2\L_n[\~V'-\~V^2]}.
\]
It is this last pair of equations that form the \bk\ transformation for
the $\PII$ hierarchy, mapping the pair $(\~V, \a_n)$ to the pair
$(V,\alpha_n)$.
Taking into account also the discrete symmetry of the $\PII$ hierarchy,
$(V,\alpha_n)
\rightarrow (-V,-\alpha_n)$, we can write our \bk\ transformation as
\be
V = \~V +\frac{2\a_n-\ep}{X-2\L_n[\ep\~V'-\~V^2]},\qquad
%\label{hbt1} \\
\alpha_n = \ep -\a_n,
\label{hbt12}
\ee
where
$\ep =\pm 1$. Using this discrete symmetry is equivalent to using the
truncation
$V=(\varphi'/\varphi)+\~V$,
that is, the truncation for the second principal family of the $\PII$
hierarchy.
We note that this \bk\ transformation is an involution,
if we iterate keeping the same value of $\ep$. We also note that it preserves
the quantity $\ep V'-V^2$; for $\~V$ a solution of (\ref{p2A}), we have
$\ep V'-V^2 = \ep \~V'-\~V^2$.

Thus we have obtained, using a modification of the truncated \p\ expansion
technique, the
auto-B\"ack\-lund transformations of Airault \cite{HA}, one for each member
of the $\PII$ hierarchy.
These could now be used to generate sequences of rational solutions for
members of
the $\PII$ hierarchy or, as we shall see later, sequences of special
integrals. We
note that an alternative approach to the construction of rational solutions for
members of the $\PII$ hierarchy can be found in \cite{KP98}.

For $n=1$, equations (\ref{hbt12}) give the \bk\ transformation (\ref{P2BTb})
for $\PII$. If $n=2$ then it gives %(recall $\ep=\pm 1$)
\be
V = \~V+\frac{2\a_2-\ep}{X-2[\ep \~V'''-2\~V
  \~V''+\~(V')^2-6\ep\~V^2\~V'+3\~V^4]},\qquad
%\label{yt1} \\
\alpha_2 = \ep -\a_2
\label{yt2}
\ee
with $\ep=\pm 1$. It is easy to verify that this is indeed a \bk\
transformation for
equation (\ref{p24}). The \bk\ transformation in the special case $n=2$ can
also
be found in \cite{Gr,Ho}.

\section{\bk\ transformation to a $\mathbf \P34$ hierarchy}

\subsection{A $\P34$ hierarchy}
Consider the $\PII$ hierarchy (\ref{p2h}), and we write
\be
Y=V'-V^2
\label{t1}
\ee
to obtain
\be
\left(\frac{\d}{\d X}+2V\right)\L_n[Y]-VX-\alpha_n=0.
\label{t2}
\ee
Equations (\ref{t1}) and (\ref{t2}) then provide a \bk\ transformation.
The result of eliminating $Y$ between these two equations is that $V$ satisfies
(\ref{p2h}). However since $\L_n[Y]$ contains $\d^{2n-2}Y/\d X^{2n-2}$, we see
that the elimination of $V$, by solving (\ref{t2}) for $V$ and substituting
into (\ref{t1}), also yields an ODE of order $2n$. That is, assuming that
$\L_n[Y]-\tfr12X\neq0$, we have an invertible transformation
\ba
&& Y = V'-V^2,\label{t3} \\
&& V = -\,\frac{1}{2\L_n[Y]-X}\left[\frac{\d}{\d
X}\left(\L_n[Y]\right)-\alpha_n\right]
\label{t4}
\ea
between (\ref{p2h}) and
\ba
&&\left(2\L_n[Y]-X\right)\frac{\d^2}{\d X^2}\left(\L_n[Y]\right)-
\left[\frac{\d}{\d X}\left(\L_n[Y]\right)\right]^2+\frac{\d}{\d
X}\left(\L_n[Y]\right)
\nonumber \\
&&\qquad +\left(2\L_n[Y]-X\right)^2Y-\alpha_n(1-\alpha_n)=0.
\label{p34h}
\ea
This sequence of ODEs (\ref{p34h}) is a hierarchy of higher order $\P34$
equations, and shall henceforth be referred to as the $\P34$ hierarchy (see
\cite{Ho} for a different derivation). Differentiating (\ref{p34h}) gives
\[
\left(\frac{\d^3}{\d X^3}+4Y\frac{\d}{\d X}+2Y'\right)\L_n[Y]-XY'-2Y=0,
\]
that is,
\be
\frac{\d}{\d X}\left(\L_{n+1}[Y]\right)-XY'-2Y=0.
\label{dp34h}
\ee
This last is just the similarity reduction of the KdV hierarchy (\ref{kdv})
obtained via
\[
U = \frac{Y(X)}{[(2n+1)t_{2n+1}]^{{2}/({2n+1})}}, \qquad
X = \frac{x}{[(2n+1)t_{2n+1}]^{{1}/({2n+1})}},
\]
which is as should be expected for a $\P34$ hierarchy.

The \bk\ transformation (\ref{t3},\ref{t4}) appears in \cite{HA}, but the
$\P34$ hierarchy
(\ref{p34h}) does not. Instead the result of eliminating $V$ was found to
be (\ref{dp34h}). In
\cite{NK97}, the result that for any solution $V$ of (\ref{p2h}), $Y$
defined by (\ref{t3}) gives
a solution of (\ref{dp34h}) was also obtained, but in the special case
$\alpha_n=\tfr12$.

For $n=1$ we obtain the \bk\ transformation
\be
Y = V'-V^2,\qquad
%\label{fa1} \\
V = -\frac{Y'-\alpha_1}{2Y-X}
\label{fa2}
\ee
between $\PII$, i.e.\ equation (\ref{p2}) with parameter $\alpha=\alpha_1$, and
\be
(2Y-X)Y''-(Y')^2+Y'+(2Y-X)^2Y-\alpha_1(1-\alpha_1)=0.
\label{p34}
\ee
This equation can be mapped to $\P34$ of \cite{Ince}. The \bk\
transformation (\ref{fa2}), for this case $n=1$, can be found in
\cite{FA} (with $(V,\alpha_1) \rightarrow (-V,-\alpha_1)$).

For $n=2$ we obtain the invertible transformation
\[
Y = V'-V^2,\qquad
%\label{g1} \\
V = -\frac{Y'''+6YY'-\alpha_2}{2(Y''+3Y^2)-X}
\]
between equation (\ref{p24}) and
\ban
&&\left[2(Y''+3Y^2)-X\right]\frac{\d^2}{\d X^2}(Y''+3Y^2) - \left[\frac{\d}{\d
X}(Y''+3Y^2)\right]^2 +\frac{\d}{\d X}(Y''+3Y^2) \\ &&\qquad
+\left[2(Y''+3Y^2)-X\right]^2Y
-\alpha_2(1-\alpha_2)=0.
\ean
This is the fourth order equation in the $\P34$ hierarchy, which we give
here for the first time.

Note that by considering what happens when the above \bk\ transformation
breaks down, we find that:
\begin{itemize}
\item[(i)] if $Y$ is a solution of $2\L_n[Y]-X=0$ then it is
also a solution of (\ref{dp34h});
\item[(ii)] from such a solution of $2\L_n[Y]-X=0$ a solution $V$ of
(\ref{p2h}) in the special
case $\alpha_n=\tfr12$ can be obtained from (\ref{t3}).
\end{itemize}

\noindent These two results can in fact be found in \cite{NK97}; the first
appears also in \cite{HA} for the special cases $n=1$ and $n=2$. Note that
the hierarchy $\L_n[Y]-(X/2)=0$, which gives rise to special integrals of
the $\PII$ hierarchy, is just the $P_{\! I}$ hierarchy.

\subsection{An alternative formulation}
Here we give an alternative description of the $\P34$ hierarchy
(\ref{p34h}). This we do by writing
down a second hierarchy of ODEs which has (\ref{p34}) as its first member,
and then giving an
invertible transformation between this second hierarchy and the hierarchy
(\ref{p34h}). Let us
consider once again the $\PII$ hierarchy (\ref{p2h}), but this time instead
of (\ref{t3}) we write
\be
Z=\L_n[V'-V^2]
\label{sbt1}
\ee
to obtain
\be
Z'+2VZ-VX-\alpha_n=0.
\label{sbt2}
\ee
Equations (\ref{sbt1}) and (\ref{sbt2}) again provide a \bk\ transformation
such that the equation satisfied by $V$ is (\ref{p2h}). Since
$\L_n[V'-V^2]$ contains
$\d^{2n-1}Z/\d X^{2n-1}$, we see that the elimination of $V$ yields an ODE
in $Z$ also
of order $2n$. That is, we have the invertible transformation
\[
Z = \L_n[V'-V^2],\qquad
V = -\frac{Z'-\alpha_n}{2Z-X}
\]
between (\ref{p2h}) and
\be
\L_n\left[\frac{(Z')^2-Z'-Z''(2Z-X)+\alpha_n(1-\alpha_n)}{(2Z-X)^2}\right]-Z=0.
\label{p34h2}
\ee
It is easy to see that for $n=1$ equation (\ref{p34h2}) reduces to (\ref{p34}),
and that for $n=2$ it gives
\ban
&&\frac{\d^2}{\d
X^2}\left[\frac{(Z')^2-Z'-Z''(2Z-X)+\alpha_n(1-\alpha_n)}{(2Z-X)^2}\right]
\\
&&\qquad
+3\left[\frac{(Z')^2-Z'-(2Z-X)Z''+\alpha_n(1-\alpha_n)}{(2Z-X)^2}\right]^2-Z
=0.
\ean

The hierarchy (\ref{p34h2}) is in fact an alternative formulation of the
$\P34$ hierarchy (\ref{p34h}). This is easily seen by
considering the \bk\ transformation
\[
Z = \L_n[Y], \qquad
Y = \frac{(Z')^2-Z'-(2Z-X)Z''+\alpha_n(1-\alpha_n)}{2Z-X},
\]
which provides an invertible transformation between (\ref{p34h2}) and
(\ref{p34h}).

\section{\bk\ transformations for the case $\mathbf \alpha_n=\tfr12$}
We now consider once again the \bk\ transformation (\ref{t3},\ref{t4}),
in the special case $\alpha_n=\tfr12$. Just because we take
$\alpha_n=\tfr12$ does not
mean that (\ref{t3},\ref{t4}) breaks down. Writing
$\L_n[Y]-\tfr12X=\delta\psi^2$, where $\delta$ is a constant,
then gives a \bk\ transformation
\ba
&& \L_n[V'-V^2]-\tfr{1}{2}X-\delta\psi^2 = 0, \label{bth1} \\
&& \psi'+V\psi = 0. \label{bth2}
\ea
Elimination of $\psi$ shows that $V$ satisfies
\be
\left(\frac{\d}{\d X}+2V\right)\L_n[V'-V^2]-VX-\tfr{1}{2}=0, \qquad{}
n=1,2,3,\ldots,
\label{dd1}
\ee
as should be expected. Conversely, elimination of $V$ shows that $\psi$
satisfies
\be
\L_n[-\psi''/\psi]-\tfr{1}{2}X-\delta\psi^2=0, \qquad{} n=1,2,3,\ldots.
\label{dd2}
\ee

For $\delta=0$ we obtain from (\ref{bth1}) the special integrals
(\ref{In1}), with
equation (\ref{dd2}) becoming (\ref{peq}). However, for $\delta\neq 0$,
what we have
is a \bk\ transformation between the two equations (\ref{dd1}) and
(\ref{dd2}), each of
which is of order $2n$. We note that the \bk\ transformation
(\ref{bth1},\ref{bth2}) can also
be obtained by considering the general integral of (\ref{se}) in the
special case $\alpha_n=\tfr12$
(this is the case where the constant of integration cannot be removed by a
simple linear
transformation $\sigma\rightarrow\sigma+c$, for some constant $c$).

Equations (\ref{dd2}) are of course related to those of the $\P34$
hierarchy (\ref{p34h}) with $\alpha_n=\tfr12$ through the \bk\ transformation
\ba
&& \L_n[Y]-\tfr{1}{2}X-\delta\psi^2 = 0, \label{cc1} \\
&& \psi''+Y\psi = 0, \label{cc2}
\ea
as is easily shown by the elimination of $Y$ and $\psi$ respectively from
(\ref{cc1},\ref{cc2}).
We note that when $\delta=0$ the system (\ref{cc1},\ref{cc2}) provides an
alternative formulation
of the special integrals (\ref{In1}). It is straightforward, using the
symmetry $(V,\alpha_n)
\rightarrow (-V, -\alpha_n)$, to give \bk\ transformations corresponding to
(\ref{bth1},\ref{bth2}) and (\ref{cc1},\ref{cc2}) in the case
$\alpha_n=-\tfr12$.

The reason for our interest in this special case $\alpha_n=\tfr12$, and the
hierarchy
(\ref{dd2}), will now become clear.

\subsection{A special auto-\bk\ transformation for $\PII$}
For $n=1$ we obtain from the above the \bk\ transformation
\ba
&& V'-V^2-\tfr{1}{2}X-\delta\psi^2 = 0, \label{p2a} \\
&& \psi'+V\psi = 0, \label{p2b}
\ea
between
\[
V''-2V^3-XV-\tfr{1}{2}=0
\]
and
\be
\psi''+\delta\psi^3+\tfr{1}{2}X\psi=0.
\label{dpsi}
\ee
% Note that in the elimination of $V$ and $\psi$ we do not need to take
% square roots
% or to divide by $\psi$.

For $\delta=0$, equation (\ref{p2a}) becomes (\ref{I11}), the first
integral $I^{(1)}_{1/2}$,
and equation (\ref{dpsi}) becomes (\ref{peq1}), giving rise to the Airy
function solutions
of $\PII$.

However, for $\delta\neq 0$, then (\ref{dpsi}) is equivalent to $\PII$ with
parameter
$\alpha_1=0$. Thus (\ref{p2a}) and (\ref{p2b}) provide a special auto-\bk\
transformation for $\PII$, mapping between solutions of $\PII$ for $\alpha_1=0$
and solutions for $\alpha_1=\tfr12$. This \bk\ transformation is not
well known though it was first written down by Gambier\footnote{We are grateful
to Chris Cosgrove for this information.} \cite{Gam}.
It is easy to see from (\ref{p2a}) and
(\ref{p2b})
that this \bk\ transformation provides a connection between the simplest
Airy function
solution of $\PII$ (for $\alpha_1=\tfr12$) and the zero solution of $\PII$
(for $\alpha_1=0$),
and thus between the Airy function solution hierarchy of $\PII$ (for
$\alpha_1=n+1/2$, with $n$
an integer) and the rational solution hierarchy of $\PII$ (for $\alpha_1$
an integer).

\subsection{A \bk\ transformation for $\mathbf \alpha_2=\tfr12$}
For $n=2$ we obtain from (\ref{bth1},\ref{bth2}) the \bk\ transformation
\ba
&&V'''-2VV''+(V')^2-6V^2V'+3V^4-\tfr{1}{2}X-\delta\psi^2 = 0, \label{n21} \\
&&\psi'+V\psi = 0, \label{n22}
\ea
between
\be
V''''-10V^2V''-10V\left(V'\right)^2+6V^5-XV-\tfr{1}{2}=0
\label{n23}
\ee
and
\[
\psi^2\psi''''-2\psi\psi'\psi'''-4\psi(\psi'')^2+2(\psi')^2\psi''+\tfr{1}{2}X
\psi^3+\delta\psi^5=0.
\]

For $\delta=0$, (\ref{n21}) gives the special integral $I^{(2)}_{1/2}$
(\ref{I21})
of (\ref{n23}). This special integral is in fact equivalent to a special case
of
Chazy Class XI ($k=3$) \cite{Chazy}, the additional non-dominant terms
being such
that it has, according to Chazy, the \p\ property. In the variables used by
Chazy ($y=-\tfr12V$, $z=-\tfr12Y$, $a=0$, $b=-\tfr14X$), this special
integral can be written
(compare with (\ref{cc1},\ref{cc2}) for $n=2$ and $\delta=0$),
\ba
&& z'' = 6z^2-\tfr{1}{4}X \label{lam1} \\
&& \psi'' = 2z\psi \label{lam2}
\ea
the second of which is a generalised Lam\'e equation, the usual elliptic
function
being replaced by a solution of the first \p\ equation $P_{\! I}$ (a suitable
rescaling of $z$ and $X$ brings (\ref{lam1}) to standard form). The special
integrals
(\ref{In1}) then give a whole hierarchy of such equations; the system
(\ref{bth1}),
(\ref{bth2}), or equivalently the system (\ref{cc1},\ref{cc2}), then
represents a
further generalisation. In each case $V$ satisfies a member of the $\PII$
hierarchy.

\section{Special integrals of the $\mathbf P_{\! I\! I}$ hierarchy}
In this section we consider the derivation of special integrals for the $\PII$
hierarchy. We do this by choosing $\ep=1$ and then using the mapping
$(V,\alpha_n)\rightarrow (-V,-\alpha_n)$ so that our \bk\ transformation
(\ref{hbt12}) becomes
\be
V = -\~V -\frac{2\a_n-1}{X-2\L_n[\~V'-\~V^2]}, \qquad
%\label{hbt1a} \\
\alpha_n = -1+\a_n,
\label{hbt2a}
\ee
Beginning with the special integrals (\ref{In1}),
\be
I^{(n)}_{1/2}\equiv \L_n[V'-V^2]-\tfr{1}{2}X=0,
\label{SI}
\ee
which define solutions of (\ref{p2h}) for $\alpha_n=\tfr12$, the \bk\
transformation (\ref{hbt2a}) then allows us to express these
special integrals in terms of solutions of (\ref{p2h}) but now for
$\alpha_n=\tfr32,\tfr52,\ldots$ (so for the first step we take
$\alpha_n=\tfr12$
and $\a_n=\tfr32$ and express (\ref{SI}) in terms of $\~V$). Special
integrals for $\alpha_n=-\tfr12,-\tfr32,-\tfr52,\ldots$ are obtained using the
discrete symmetry $(V,\alpha_n)\rightarrow (-V,-\alpha_n)$. In this way we are
able to give for each member of the $\PII$ hierarchy (\ref{p2h}) a sequence
of special integrals, one for each half odd integer value of $\alpha_n$.

For $n=1$, the \bk\ transformation (\ref{hbt2a}), is
\[
V = -\~V -\frac{2\a_n-1}{X-2(\~V'-\~V^2)}, \qquad
%\label{hbt1c} \\
\alpha_n = -1+\a_n.
%\label{hbt2d}
\]
Beginning with the special integral $I^{(1)}_{1/2}$ (\ref{I11}), that is,
\[
I^{(1)}_{1/2} \equiv V'-V^2-\tfr{1}{2}X = 0,
\]
we obtain immediately the corresponding special integral for
$\alpha_1=-\tfr12$,
\[
I^{(1)}_{-1/2} \equiv V'+V^2+\tfr{1}{2}X = 0,
\]
and then using the above procedure we obtain
\ban
I^{(1)}_{\pm 3/2} & \equiv & (V')^3 \mp\left( V^2 +\tfr{1}{2}X \right) (V')^2
-\left( V^4 +XV^2 \pm 4V +\tfr{1}{4}X^2 \right) V' \\ & &
\pm V^6 \pm\tfr{3}{2}XV^4 +4V^3 \pm\tfr{3}{4}X^2V^2 +2XV \pm\tfr{1}{8}X^3
\pm 2 =0,
\ean
and
\ban
I^{(1)}_{\pm 5/2} & \equiv & (V')^5 \mp\left( V^2 +\frac{1}{2}X \right)(V')^4
-\left( 2V^4 +2XV^2 \pm 12V +\frac{1}{2}X^2 \right)(V')^3
 \\ & &
+\left( \pm 2V^6 \pm 3XV^4 +12V^3
\pm\frac{3}{2}X^2V^2 +6XV \pm\frac{1}{4}X^3 \pm 6 \right) (V')^2
 \\ & &
+\left( V^8 +2XV^6
\pm 12V^5 +\frac{3}{2}X^2V^4 \pm 12XV^3 +\frac{1}{2}X^3V^2 +36V^2 \right.
 \\ & &
\left. \pm 3X^2V +\frac{1}{16}X^4
+2X \right) V' \mp V^{10} \mp\frac{5}{2}XV^8 -12V^7 \mp\frac{5}{2}X^2V^6
 \\ & &
-18XV^5 \mp \frac{5}{4}X^3V^4 \mp 42V^4 -9X^2V^3 \mp\frac{5}{16}X^4V^2 \mp
26XV^2
 \\ & &
-\frac{3}{2}X^3V -32V \mp\frac{1}{32}X^5 \mp\frac{5}{2}X^2.
\ean

It was shown in \cite{AMM} that in this case of $\PII$ itself, these
special integrals
$I^{(1)}_{\pm\alpha_1}$, for $\alpha_1=n+1/2$ with $n$ an integer, are the
only possible
special integrals of polynomial type, and that they satisfy the equation
\[
\left(\frac{\d}{\d X}\pm 2V\right)I^{(1)}_{\pm\alpha_1}=0 \quad ({\rm modulo~}
\PII^{(1)}[V,\pm\alpha_1]=0),
\]
as can be easily checked for those examples given above.

For $n=2$ we take our \bk\ transformation in the form
\[
V = -\~V-\frac{2\a_2-1}{X-2[\~V'''-2\~V \~V''+\~(V')^2-6\~V^2\~V'+3\~V^4]},
\qquad
\alpha_2 = -1+\a_2,
\]
and, beginning with $I^{(2)}_{1/2}$ (\ref{I21}),
\[
I^{(2)}_{1/2}\equiv V''' -2VV'' +(V')^2-6V^2V' +3V^4 -\tfr{1}{2}X=0,
\]
we obtain immediately
\[
I^{(2)}_{-1/2}\equiv V''' +2VV'' -(V')^2 -6V^2V' -3V^4 +\tfr{1}{2}X=0.
\]
Then, by following the above procedure, we obtain a sequence of special
integrals,
one for each half odd integer value of $\alpha_2$. For $\alpha_2=\pm \tfr32$ we
get
\ban
I^{(2)}_{\pm 3/2} & \equiv & (V''')^3 -\left[ \pm 2VV'' \mp (V')^2 +18V^2V'
\mp 3V^4 \pm\frac{1}{2}X \right](V''')^2 \\ & &
-\left[ 4V^2(V'')^2 -4V(V')^2V'' \mp 24V^3V'V'' -12V^5V''+2XVV''
\phantom{\frac{1}{2}} \right. \\ & &
\pm 4V'' +(V')^4 \pm 12V^2(V')^3 -102V^4(V')^2 -X(V')^2 \pm 36V^6V' \\ & &
\left. \mp 6XV^2V' +9V^8 -3XV^4 \mp 8V^3
+\frac{1}{4}X^2 \right] V''' \pm 8V^3(V'')^3 \\ & &
-\left[ \pm 12V^2(V')^2 -24V^4V' \pm 36V^6 \mp 6XV^2
-8V \right] (V'')^2 \\ & &
+\left[ \pm 6V(V')^4 -24V^3(V')^3 \mp 36V^5(V')^2 \mp 6XV(V')^2 -4(V')^2
\phantom{\frac{1}{2}} \right. \\ & &
-72V^7V' +12XV^3V' \pm 24V^2V' \pm 54V^9 \mp 18XV^5 -28V^4 \\ & &
\left. \pm\frac{3}{2}X^2V +2X \right] V''
\mp (V')^6 +6V^2(V')^5 \pm 27V^4(V')^4 \pm\frac{3}{2}X(V')^4 \\ & &
-180V^6(V')^3 -6XV^2(V')^3 \pm 81V^8(V')^2 \mp 9XV^4(V')^2 +8V^3(V')^2 \\ & &
\mp\frac{3}{4}X^2(V')^2 +54V^{10}V' -18XV^6V' \mp 48V^5V'
+\frac{3}{2}X^2V^2V' +4V' \\ & &
\mp 27V^{12} \pm\frac{27}{2}XV^8 +24V^7 \mp\frac{9}{4}X^2V^4 -4XV^3 \mp 4V^2
\pm\frac{1}{8}X^3 =0. \\ & &
\ean

Unfortunately the length of successive special integrals prevents us from
giving them explicitly here. For example, $I^{(2)}_{\pm 5/2}$ contains
$318$ terms and the
first new special integral obtained at $n=3$, $I^{(3)}_{\pm 3/2}$, contains
$281$ terms. We have checked that
\[
\left(\frac{\d}{\d X}\pm 2V\right)I^{(2)}_{\pm\alpha_2}=0 \quad ({\rm modulo~}
\PII^{(2)}[V,\pm\alpha_2]=0)
\]
for $\alpha_2=\tfr12$, $\alpha_2=\tfr32$ and $\alpha_2=\tfr52$, and also that
\[
\left(\frac{\d}{\d X}\pm 2V\right)I^{(3)}_{\pm\alpha_3}=0 \quad ({\rm modulo~}
\PII^{(3)}[V,\pm\alpha_3]=0)
\]
for $\alpha_3=\tfr12$ and $\alpha_3=\tfr32$. In general, the special integral
$I^{(n)}_{\pm\alpha_n}$, $\alpha_n=\tfr12,\tfr32,\ldots$, is a polynomial of
degree $2\alpha_n$ in $\d^{2n-1}V/\d X^{2n-1}$ and satisfies
\[
\left(\frac{\d}{\d X}\pm 2V\right)I^{(n)}_{\pm\alpha_n}=0 \quad ({\rm modulo~}
\PII^{(n)}[V,\pm\alpha_n]=0).
\]

We note that although we have given a method of constructing special integrals
for higher members of the $\PII$ hierarchy for $\alpha_n$ any half odd integer,
the question of completeness, as studied in \cite{AMM} for $\PII$, remains
open.

\section{Conclusions}
We have shown how to modify the truncation procedure in \p\ analysis so
that it may be used to find auto-\bk\ transformations of ODEs. Our main
observations are that we have to take into account the possibility of changes
in the values of parameters in the ODE, and that this then implies that we
cannot set all coefficients of $\varphi$ equal to zero independently. The main
difference between our approach and that for PDEs is that we eliminate the
singular manifold from the truncated expansion, which we then rewrite in terms
of a second solution of the ODE. This approach is further extended in
\cite{GJP}.

We have also given a variety of other \bk\ transformations, including one
between the $\PII$ hierarchy and a $\P34$ hierarchy, and Gambier's special
auto-\bk\ transformation for $\PII$ itself, this last relating $\PII$ with
$\alpha_1=0$ to $\PII$ with $\alpha_1=\tfr12$. In addition we have also
considered the
construction of special integrals for the $\PII$ hierarchy, and have shown
how to
construct such integrals for $\alpha_n$ any half odd integer. Special integrals
have not previously been given for the higher members of this hierarchy.

\vspace{-0.3truecm}
\section*{Acknowledgements}
Andrew Pickering is very grateful to Nalini Joshi for her invitation to
make an extended visit to
the University of Adelaide. The research in this paper of NJ and AP was
supported by the Australian
Research Council.
AP is grateful to PAC for the use of Mathematica in Canterbury.

%\newpage
\vspace{-0.3truecm}
\section*{References}
\def\refjl#1#2#3#4#5#6{{\rm #1\ #2\ #3\ {\frenchspacing\em#4}\ {\bf #5}\ #6}}
\def\refbk#1#2#3#4{{\rm#1}\ #2\ {\sl#3}\ (#4)}

\end{document}